# Assessing Visualization Techniques for the Search Process in Digital Libraries

**Wilko van Hoek[1], Philipp Mayr**
GESIS - Leibniz-Institute for the Social Sciences
Unter Sachsenhausen 6-8
50667 Cologne, Germany
{Wilko.vanHoek, Philipp.Mayr}@gesis.org

## Abstract

In this paper we present an overview of several visualization techniques to support the search process in Digital Libraries (DLs). The search process typically can be separated into three major phases: query formulation and refinement, browsing through result lists and viewing and interacting with documents and their properties. We discuss a selection of popular visualization techniques that have been developed for the different phases to support the user during the search process. Along prototypes based on the different techniques we show how the approaches have been implemented. Although various visualizations have been developed in prototypical systems very few of these approaches have been adapted into today's DLs. We conclude that this is most likely due to the fact that most systems are not evaluated intensely in real-life scenarios with real information seekers and that results of the interesting visualization techniques are often not comparable. We can say that many of the assessed systems did not properly address the information need of current users.

## Introduction

A Digital Library (DL) can be defined as:

> a focused collection of digital objects, including text, video, and audio, along with methods for access and retrieval, and for selection, organization, and maintenance of the collection. (Witten, Bainbridge, & Nichols, 2009)

---

[1] Corresponding author, Wilko.vanHoek@gesis.org

There are different definitions of DLs, however, they all agree on at least one point, namely that DLs hold digital objects which are accessible by users (Fox, Gonçalves, & Shen, 2012). Therefore, besides the challenges of "selection, organization, and maintenance of the collection" (Witten et al., 2009), "access and retrieval" of the stored objects or rather documents is an important task in DLs. As the documents stored in a DL usually do not change over the time, we even believe that this fact is one of the most crucial ones in DLs. To provide a search interface is the way to make content for users accessible.

Developing and improving search interfaces for DLs has always to align with the type of documents stored in the DL. Textual data is one of the main document types that are stored in today's scientific DLs. The usual paradigm for search interfaces is "type-keywords-in-entry-form, view-results-in-a-vertical-list" (M. Hearst, 2009). Hence, an advantage of textual content is that it is searchable. Nevertheless, there is also a disadvantage. In the fields of Human-Computer Interaction (HCI) and Information Visualization (InfoVis) it is argued that visualizations in search interfaces pose a large potential to improve the user's experience and performance in satisfying his information needs. On the contrary, visualizing textual data is a very difficult task (M. Hearst, 2009), leading to a conflict that has not yet been satisfyingly solved.

We can look back at roughly 20 years of research in visualizations for search interfaces in DLs. Many different proposals have been made and many studies have been conducted to assess search interfaces. Until now only minor contributions have been adapted to today's DLs. More complex ideas on implementing highly visual search interfaces were often discontinued.

In this paper we want to give a brief overview on some research that has been made in the field of search interfaces for textual DLs. In addition we want to discuss why most of the results of the research have not been adapted into today's DLs. To answer these questions we will look on a set of examples of techniques and systems that have been developed and assessed during the last two decades. Due to the size limits of this paper, the list of example techniques cannot and is not intended to be complete.

For a better understanding and comparability, all techniques will be introduced according to the following structure. First we give a short explanation of the idea and the functionality of the technique, as well as an example figure. Then we will give a summarized history on the development of the technique and point to existing variants alongside with the systems that use the technique. We will also take a look on how far the techniques have been evaluated.

We will organize this paper according to our own definition of a search process. We find the search process to be separable into three main interaction parts. The first part consists of specifying a query that has to be processed by a system, the second part covers the resulting view that the system offers to the user based on the query. The third part is the document view selected by the user in which detailed information about a document is presented.

This paper is structured as follows. First we give an introduction on the difficulties that arise when visualizing textual data. Then we will introduce three sections about visualization techniques. The first on query specification and refinement, the second on result views and the third on document view. At the end we will discuss the questions stated above and try to draw conclusions on what further steps should be taken and what aims should be defined for visualizations to aid the search process in the future.

## Visualizing Textual Data for Digital Libraries

The main goal when visualizing data in DLs is to aid the user during his retrieval-task. Visualizations are supposed to give overviews on or show connections between the content or to support the user with additional relevant information. Before developing visualizations it is important to understand what kind of data has to be visualized. In general DLs two kinds of data are stored, the metadata, describing the stored content, and the content itself. While the content stored in a DL may be videos, books, audio files as well as pictures, the metadata, which is stored along the content, is classified differently.(M. Hearst, 2009)

A common way to classify data is to distinguish between quantitative and categorical data:

> Quantitative data is defined as data upon which arithmetic operations can be made (integers, real numbers). Categorical data can be further subdivided into interval, ordinal, nominal, and hierarchical. Interval data is essentially quantitative data that has been discretized and made into ordered data (e.g., time is converted into months, quarters, and years). Ordinal data is data that can be placed in an order, but the differences among the values cannot be directly measured (e.g., hot, warm, cold, or first, second, third). Nominal data includes names of people and locations, and text. Finally, hierarchical data is nominal data arranged into subsuming groups. (M. Hearst, 2009)

To differ between these different kinds of data is very important for visualizations. Quantitative data can be visualized easily and reasonably with known techniques like graphs or charts. In Figure 1a, the numbers of documents per year are displayed for a set of search results. It can easily be seen that most documents retrieved were published around the year of 2010. Such visualization is helpful and can be created directly from the metadata by aggregating the number of documents in groups.[2]

Visualizing nominal data in a reasonable fashion is much more difficult.

---

[2] E.g. on http://www.gopubmed.org statistics for a search can be display directly along the results.

If we aggregate the number of documents per author for the same result set (cf. Figure 1b), the same distribution does not hold the same kind of information. A trend interpretation regarding the close neighbourhood of the peak in this case cannot be done as the fact that author C is close to author B and author C is a coincidence which is based on the alphabetical sortation of the author names. A much more informative visualization regarding authors is to visualize e.g. a co-authorship network (cf. Figure 2). In these networks central authors can be identified and thus more, conclusions be drawn. However, specialized visualizations like network graphs require more effort in calculation of relations and implementation of visualization services. Finding suitable visualizations that can be generated in runtime is one of the main challenges in the field of information visualization and thus, in DL visualization. In the past decades a variety of different approaches on visual support in the search process has been developed. Some of them propose novel prototypes that break with paradigms like result lists of document surrogates or strict text-based query specification. Others are embedded into existing systems to improve the user's search experience and show only minor changes to the common interface paradigm.

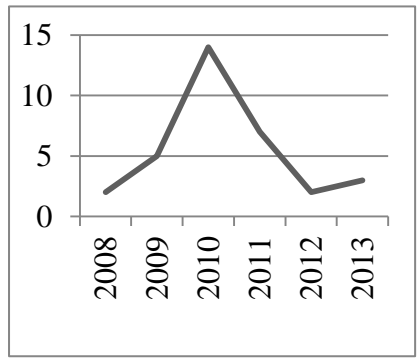
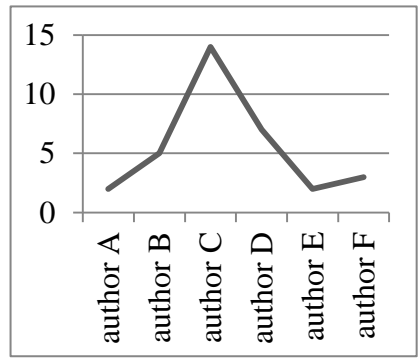

Figure 1a: Number of documents per author for a search result set.

Figure 1b: Number of documents per year for a search result set.

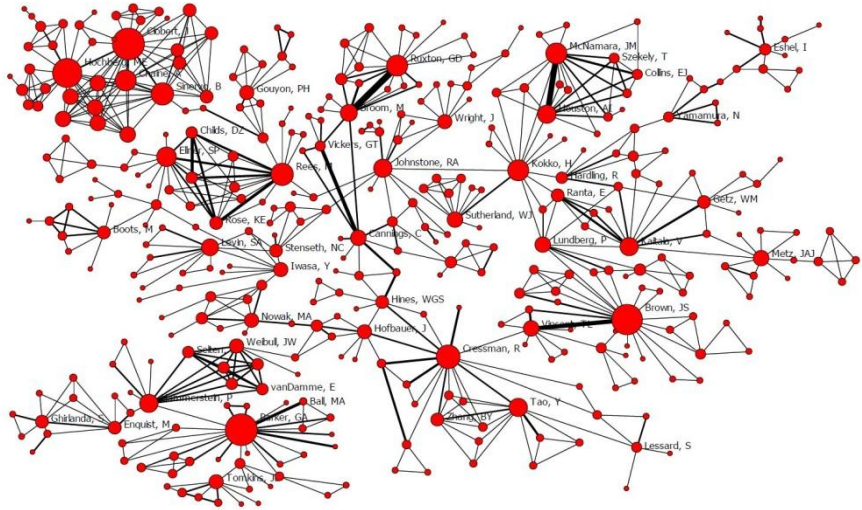

Figure 2: Co-authorship network (Morel et. al. 2009)

## Visual Support for Query Refinement

In this section we will describe a set of techniques that have been developed to support the user of a DL during the query formulation task. We define this task to be the first step of the search process. It is the starting point for every search and subsequent searches and a central point for the interaction between user and system.

### Boolean Query Specification

The search process in today's DLs starts with entering a query that consists of one or more query terms. Generally, the query can be specified by combining terms using Boolean operators such as AND, OR and NOT. As the result set relies on the query as the only explicit information that the user provides to the system about his information needs, using Boolean operators can be crucial to the task of retrieving and thus, finding the right documents.

To support the user in specifying his query, systems have been developed to visualize the query and allow him to specify the Boolean operators he wants to use or rather whether he wants intersection, union or difference in relation to the query terms. This can be done by using Venn diagrams. Figure 3 shows the user interface of the VQuery system (Jones, 1998). The user can enter query terms which are then added to the left pane where an area can be modified to determine which terms are used for the query. By moving a circle on others the user can connect the terms they represent and thus, specify what Boolean expression is to be generated. In the lower left corner the resulting query is displayed in verbal form. The retrieved document titles are presented in the right pane.

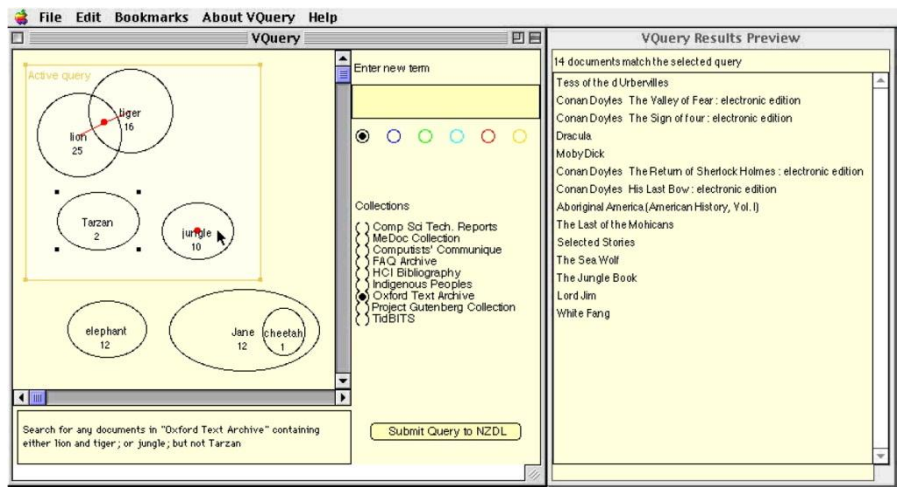

Figure 3: Query Specification with VQuery (Jones, 1998). Each query is represented as a circle that can be moved by the user. Positions and overlaps of query terms imply a Boolean query that can then be executed.

Visual query specification has been introduced to database management systems in the early 1980s. The idea to adapt this technique for DLs took until the late 1990s. The usability study conducted in the VQuery context (Jones, McInnes, & Staveley, 1999) showed good results, as well as the study by Hertzum and Frøkjær. 1996 that showed significant speed-ups by using Venn diagrams for query specification along with raising correctness of the queries. Despite these promising results visual query specification has not been adapted in DLs. This could be due to the higher effort that has to be put into interface design and the progress that has been made in fuzzy search techniques. In 2011 visual query specification has been picked up again within the INVISQUE system (Wong, Chen, Kodagoda, Rooney, & Xu, 2011). Figure 4 shows two query result sets in the INVISQUE system. The sets can be merged by clicking and holding one search and dragging it onto another. In this way the division of query specification and results has been surmounted.

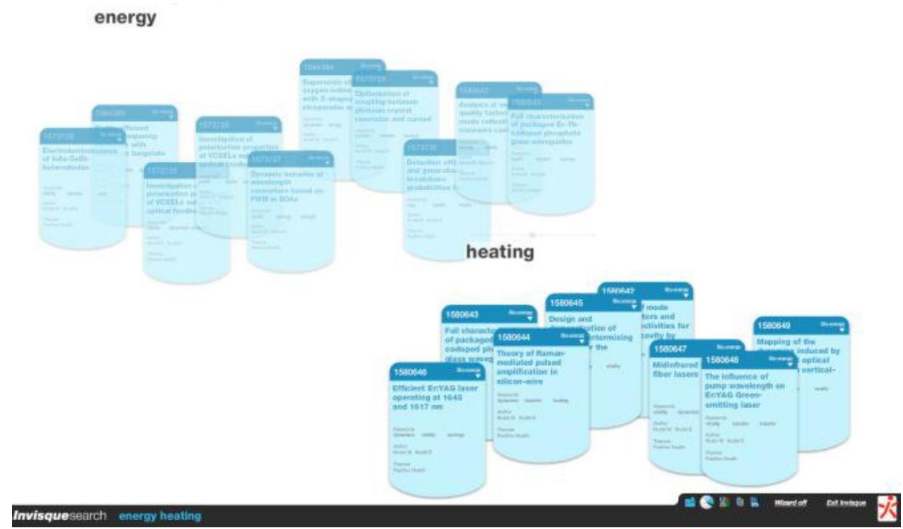

Figure 4: The INVISQUE system (Wong, Chen, Kodagoda, Rooney, & Xu, 2011) displaying query results for two queries *energy* and *heating*.

Query Term Suggestion

Term Suggestion services are an important aid in the task of formulating a query that represents a user's information need. Based on the query terms entered by the user, a service supplies the user with further query terms which he might want to use. The suggestions in DLs can usually be terms that are either close to the query terms, relevant to the query terms, terms other users have used or terms that contain the user's query terms. In general the suggestions are presented in two different ways. Either they pop up as a list while a user types in a query or they are displayed as a term cloud along with the search results after the query has been processed.

Term Suggestion is not exclusively used in DLs. In virtually all areas where queries are formulated to interact with a system, term suggestions are provided to improve the user's accuracy and swiftness. Nevertheless in each application term suggestion is used differently and the suggestions are based on other data. In DLs many effort has been made to offer terms based on metadata. General DLs often use thesauri to group documents thematically. A user who is inexperienced with a DL may not know much about the used the-sauri which makes it difficult for him to benefit from it. To overcome this problem term suggestion based on co-occurrences has been developed. An early proposal was made in 1996 (Schatz, Johnson, Cochrane, & Chen, 1996). Studies have shown that term suggestions are helpful and often used by users, but have to be domain-specific in order to be most beneficial (Hienert, Schaer, Schaible, & Mayr, 2011).

Overall query term suggestion is a key feature in DLs, but it can be argued whether it is a visual mean of helping the user. Besides, term clouds suggestions are presented as lists which cannot be seen as a complex visual experience. We argue that, considering the usefulness of term suggestion services, more research should be carried out to develop more visual ways of presenting term suggestions. Until now a user is supported with an alternative term but often not with an explanation why it is presented to him. Maybe it is possible to visually explain why a certain term is suggested and which connection exists to the user's term or even how the suggested term influences his result view.

## Visual Support in the Result View

The *de facto* standard of presenting results retrieved for a specific query in a DL is a list. As intuitive and simple as this approach is, it is argued whether it is the best way to display results as it makes minimal use of visualization techniques. In this section we will describe a set of techniques that have been developed to visually support the user of a DL during the task of browsing through query results.

### Query Highlighting

As we said before, query results in DLs typically are presented as lists. For every result a preview is created that consists of a subset of the document metadata like title, authors, journal and publication date. Based on the presented information a user has do decide which documents appeal to him and which he will ignore. A good way to support the user in this task is to enhance the document preview visually. The main problem here is to know what kind of information the user would want to be enhanced. As the query is the only explicit information that the user provides to a system, some ways have been proposed to visualize the occurrences of the query terms within the result document. A most common technique used is query highlighting.

**Visualization in digital libraries**
E Bertini, T Catarci, L Di Bello, S Kimani - From Integrated Publication and ..., 2005 - Springer
In an age characterized by tremendous technological breakthroughs, the world is witnessing overwhelming quantities and types of information. **Digital Libraries** (DLs) are a result of these breakthroughs, but they have not been spared by the challenges resulting from ...
Cited by 7   Related articles   All 8 versions   Cite

Figure 5: Document preview in Google Scholar for the query *visualization digital libraries*.

For query highlighting query terms are boldfaced in the document preview. The user can identify in which metadata fields the query terms are contained, he can get an idea of why a certain document has been retrieved and can determine whether the co-occurrence is what he meant or a random coincidence. In textual DLs this approach often is expanded from metadata to the content itself. In addition to metadata field text parts are extracted where query terms occur and are displayed in the document preview. The query terms are boldfaced and a small set of surrounding words is extracted as well, so that the context is understandable. Figure 5 shows a document preview with query highlighting in Google Scholar[3].

2-Dimensional Result Views

When it comes to query results the main paradigm is to present them in a vertical list (M. Hearst, 2009). A list of documents is a 1-dimensional object where documents can only be compared regarding one level of information (e.g. sorted by date, relevance). Thus, a variety of ideas has been proposed to overcome the list paradigm and to make use of the two dimensions a monitor can display.

We divided the different approaches into two main categories. The first are map or cluster-based techniques and the second are grid or rather table-based techniques. Both categories have in common that the underlining idea is to make use of two dimensions to display the relation between documents alongside more than one dimension. But they differ in the concept of how to derive a visualization to do so.

Map or cluster-based result lists are supposed to provide an overview over a large collection of documents. Therefore documents are displayed on a 2-dimensional pane where the position of documents to each other represents the relatedness. The relatedness is based on clustering where the multidimensionality of documents and their metadata is reduced to a 2-dimensional value. Due to the large amount of documents that are displayed in such a way they are represented by points or icons to reduce the complexity of the interface. Detailed information can be accessed upon zooming in on certain clusters or documents. Displaying results as points looks similar to star maps and thus, in some systems the idea of a map has been explicitly used and the clusters are represented as countries or islands. Figure 6 shows a cluster-based result view where clusters are presented as islands (K. Andrews, Gutl, Moser, Sabol, & Lackner, 2001).

Displaying query results in 2-dimensional clusters or as maps has been proposed since the early 1990s. One of the earliest systems is the BEAD sys-tem (Chalmers & Chitson, 1992), where the greatest amount of work had to be invested into the systems architecture and into solving problems that aroused from the clustering. An evaluation of the system was planned but results have never been published. The System InfoSky (Keith Andrews et al., 2002) introduced in 2002 combined document relatedness and hierarchical structures (e.g. collection structure, classification information).

____________________
[3] http://scholar.google.com

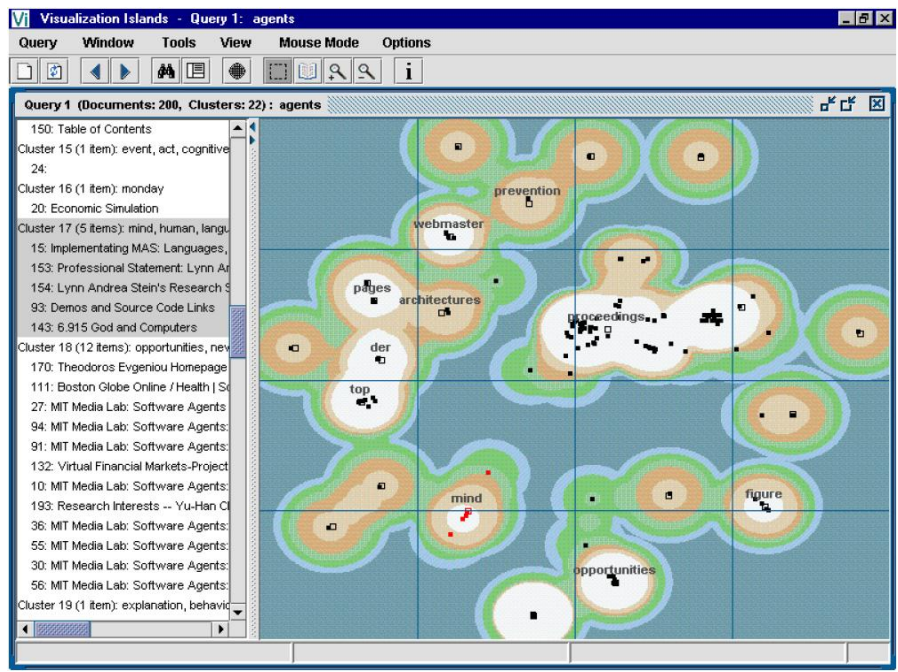

Figure 6: Document clusters displayed as islands in xFind (K. Andrews, Gutl, Moser, Sabol, & Lackner, 2001).

An initial study was conducted in which the time was measured a user needs to complete different retrieval tasks in a hierarchical tree browsing and the InfoSky system. The study showed that users needed significantly more time using the InfoSky system. A qualitative evaluation suggested that this was due to some implementation flaws and the lack of familiarity with the new system. In 2004 the authors published a second study (Granitzer, Kienreich, Sabol, Andrews, & Klieber, 2004) with a redesigned version of InfoSky. In the redesigned version of the system, a hierarchical tree browser and a star map showing clusters were shown. This combination performed better, but was still not the best choice for all tasks. The system was able to support a good overview and thus, better orientation in the vast amount of documents. However, the tree browser view performed better with users fa-miliar with the corpus. It was assumed that this was owned to the experience users had with tree-based browsing techniques.

Another way of visualizing the results in a 2-dimensional fashion is to use a grid or rather a table. Documents are arranged along their metadata information. For example all documents retrieved for a user's query are displayed row-wise by author names and column-wise by publication date. In this way the user can easily explore an author's publication list and compare it to another author's list in respect to a certain topic (cf. Figure 7). The user can change the metadata information which is used for the rows or columns, which makes it comfortable to rearrange the documents and shows their relations along other dimensions (e.g. author–topic, topic–journal). The main advantage of a table-based result list is that it relies only on the metadata. No clustering has to be calculated beforehand.

As well as the clustering approach, this method has the ability of generating a sufficient overview on large collections. To display such large result lists the documents have to be reduced to icons and numbers representing how many documents can be retrieved when a user decides to zoom in to a certain cell.

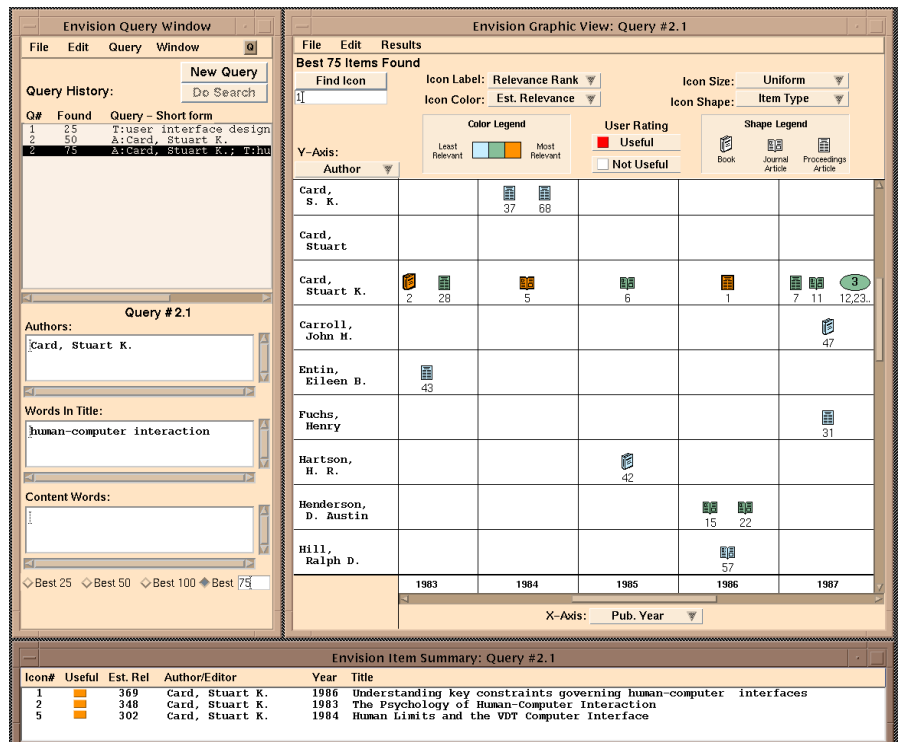

Figure 7: Results in the Envision system (Fox et al., 1993). Query term is *human-computer interaction* with author *Card, Stuart K*.

An early system using a table-based approach is the Envision system (Fox et al., 1993). In 2000 the GRIDL system (Shneiderman, Feldman, Rose, & Grau, 2000) was proposed, following the same basic idea as Envision but focussing on overcoming the problem of overcrowded rows, columns or cells. With different solutions such as tool tips or further hierarchical grouping they tried to make rows, columns or overloaded with documents more accessible. Two qualitative studies on the system's usability showed that it comprises advantages for some users and in corpora where no hierarchies exist, as it is the case in some DLs. However the users' familiarity and preferences biased the results and no proper comparisons to other systems could be made.

In 2011 a new effort was made for a table-based result view. In the VIDL system (Kim, Scott, & Kim, 2011) documents are displayed as circles whose size represents the number of pages in relation to the other results (cf. Figure 8). Through different colours the system visualizes the amount of reviews that have been written about the book and their average rating on it. In addition to the table-based result view, the system consisted of an alternative document view which we will discuss later. The usability study that was conducted showed that the result list view was assessed positively and preferred over a text-based system by 53 % of the participants. However, it was not mentioned how many participants took part in the tests. Also the system was tested based on a corpus of books and thus, a relatively small collection. It would have to be re-assessed with respect to how far the authors' approach can be utilized when applied to larger corpora.

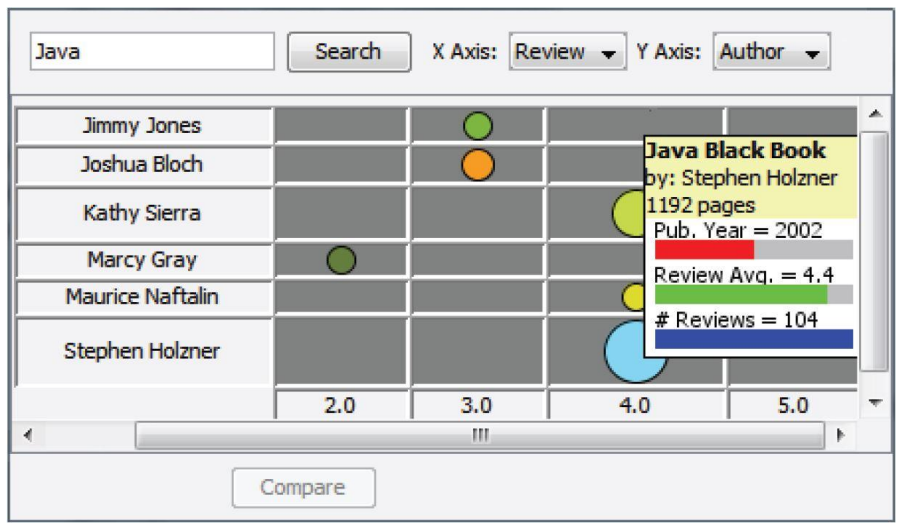

Figure 8: Search results in the VIDL system (Kim, Scott, & Kim, 2011) for the query *java*. Circle sizes indicate document sizes.

3-Dimensional Result Views

Displaying search results in a 3-dimensional space is a logical next step after developing 2-dimensional systems. Due to the fact that DLs are usually accessed by users with 2-dimensional displays, a 3-dimensional interface is not easily created. While we were able to separate 2-dimensional efforts roughly into separate categories based on the principal visualization idea, we were not able to do the same for 3-dimensional systems.

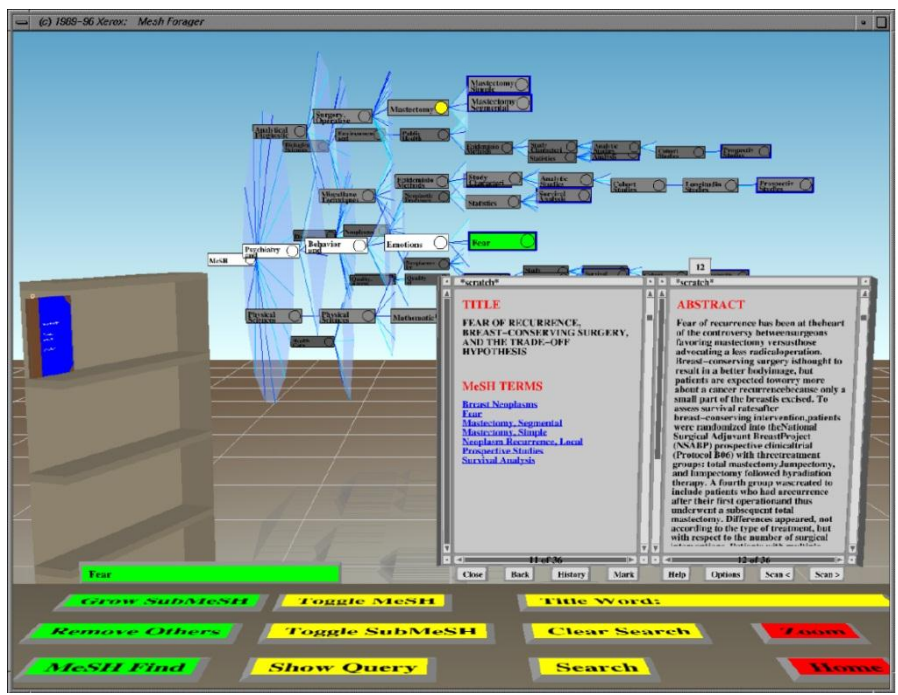

Figure 9: Search interface of the Cat-A-Cone system (M. A. Hearst & Karadi, 1997).

An early 3-dimensional system is the Cat-A-Cone system (M. A. Hearst & Karadi, 1997). The search interface is a 3-dimensional room giving the user a first person like view. Figure 9 shows the systems search interface. Apparently most elements are not 3-dimensional or do not bear more information than a 2-dimensional version would do. The document view is strictly 2-dimensional and the search controls only appear spatially. Also the bookshelf on the left side, which acts as a watch list, does not carry more information than a list. The central 3-dimensional room implemented in the system is the category visualization which allows to display category hierarchies using Cone Trees (Robertson, Mackinlay, & Card, 1991).

In the SLIS Document Space (Börner, Feng, & McMahon, 2002) a different approach to create 3-dimensional interfaces for DLs are to be found (cf. Figure 10). Here users can walk through a virtual space where websites and documents are presented as thumbnails standing on the ground. The user is represented by his avatar, a model of a human being. The idea is that users can meet on the virtual pane and chat or show their watch lists to other users. The third dimension is designed to improve interaction of users and not to enhance the search experience by adding or displaying more or more detailed information.

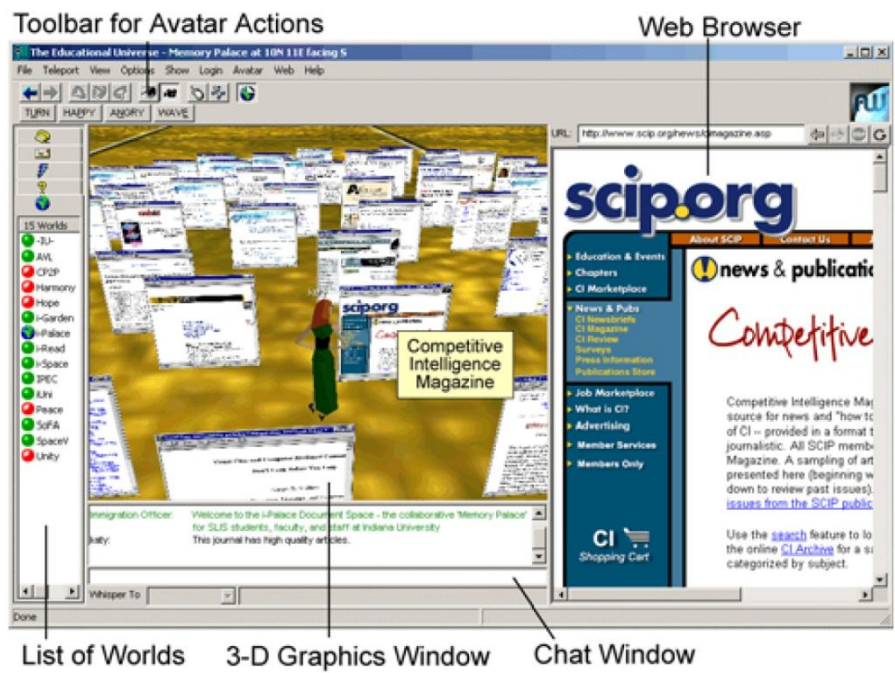

Figure 10: The SLIS Document Space (Börner, Feng, & McMahon, 2002).

The most evaluated and advanced 3-dimensional search interface seems to be the NIRVE system (Cugini, Laskowski, & Sebrechts, n.d.). For this system different concepts of displaying search results in 3-dimensions have been developed and evaluated. In the system, the user gives query terms which can be assigned to concepts. In this way the complexity of the later result view is reduced and the terms are associated in a topical way. One of the result views is a globe, where clusters of documents are displayed as boxes emerging from the globe (cf. Figure 11). The thickness of a box represents the number of documents in the cluster. The documents are clustered according to concepts defined for the query terms. Documents containing query terms belonging to a concept are assigned to that concept. Documents with the same combination of concepts build a document cluster. Starting at the south pole up to the north clusters are displayed on a latitude according to the number of concepts that they contain. Documents with one concept are displayed near the south pole, documents containing all concepts are situated near the north pole. An alphabet around the equator makes it easier to relocate a cluster. Lines connect clusters to indicate that they shared the same concept combination while one cluster possesses one more concept.

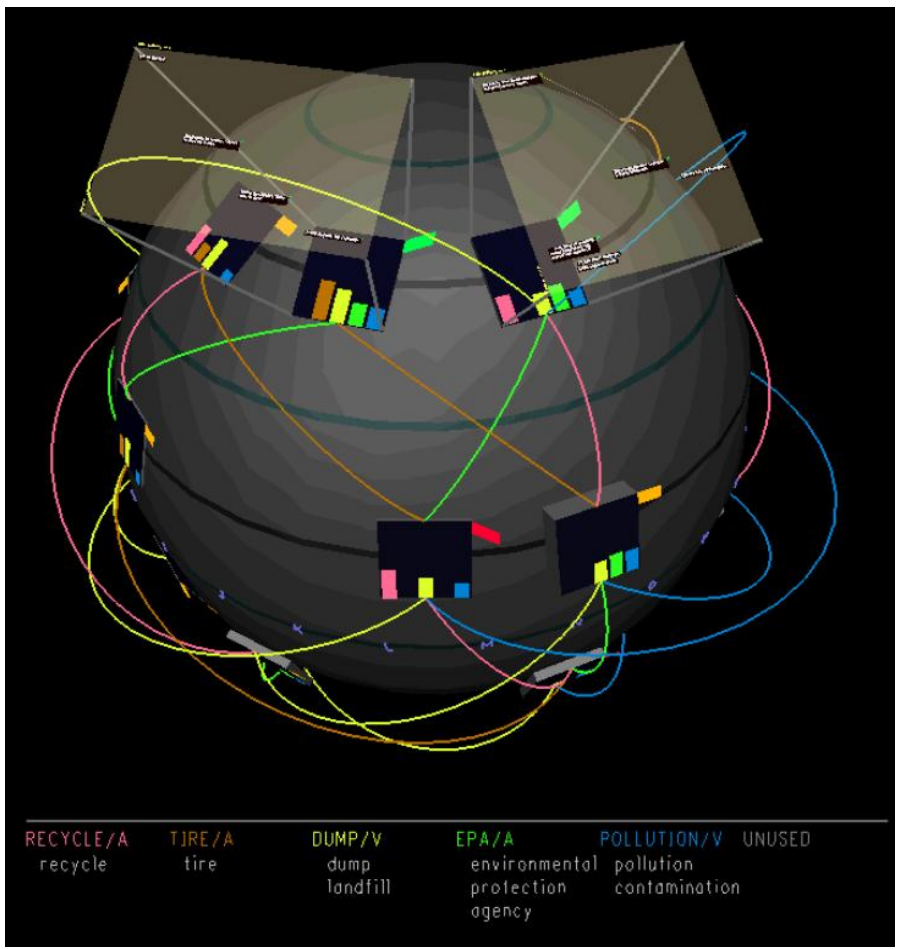

Figure 11: The Globe 3-dimensional result view in the NIRVE system (http://zing.ncsl.nist.gov/cugini/uicd/nirve-paper.html).

To assess the usability and effect of a 3-dimensional system and to compare it to a 2-dimensional interface, a study was conducted by Sebrechts, Cugini, Laskowski, Vasilakis, & Miller, 1999. Users had to fulfil tasks on a traditional text-based interface, a 2-dimensional version of the globe-based view and the globe-based view of the NIRVE system. The study showed that on most tasks users performed significantly better with a text-based interface. However, through the study, a progression could be detected. The user performances with the other two systems improved over the time, showing an increasing familiarity with the systems and their functionality.

Based on this observation a closer examination of the study revealed two main findings. Firstly, the experience users have with a system is crucial to their performance. The fact that most users are familiar with text-based systems biases results drastically. Thus, 3-dimensional systems would have to be evaluated with long term studies. Secondly, the comparability of study results is not yet given and a common study design needs to be developed.

## Visual Support in the Document View

The next step while looking through result lists is to take a closer look at certain documents. In this step the user expects to be supplied with more detailed information about the document or even to be able to access the document itself. In this section we will introduce some concepts to visualize detailed information about documents and to improve accessibility of information within a document.

### Thumbnails

Thumbnails are pictures representing a document and in contrast to icons, they are not part of a fixed set of pictures but are derived from the content of a document. They can be small images of the document itself or smaller versions of figures that appear in the document. Figure 12 shows figure-based thumbnails. Thumbnails have been evaluated and developed in the context of web search. Studies showed that thumbnails can increase a user's speed (Woodruff, Faulring, Rosenholtz, Morrsion, & Pirolli, 2001) and can be very helpful in mobile devices where typically small displays are used (Lam & Baudisch, 2005).

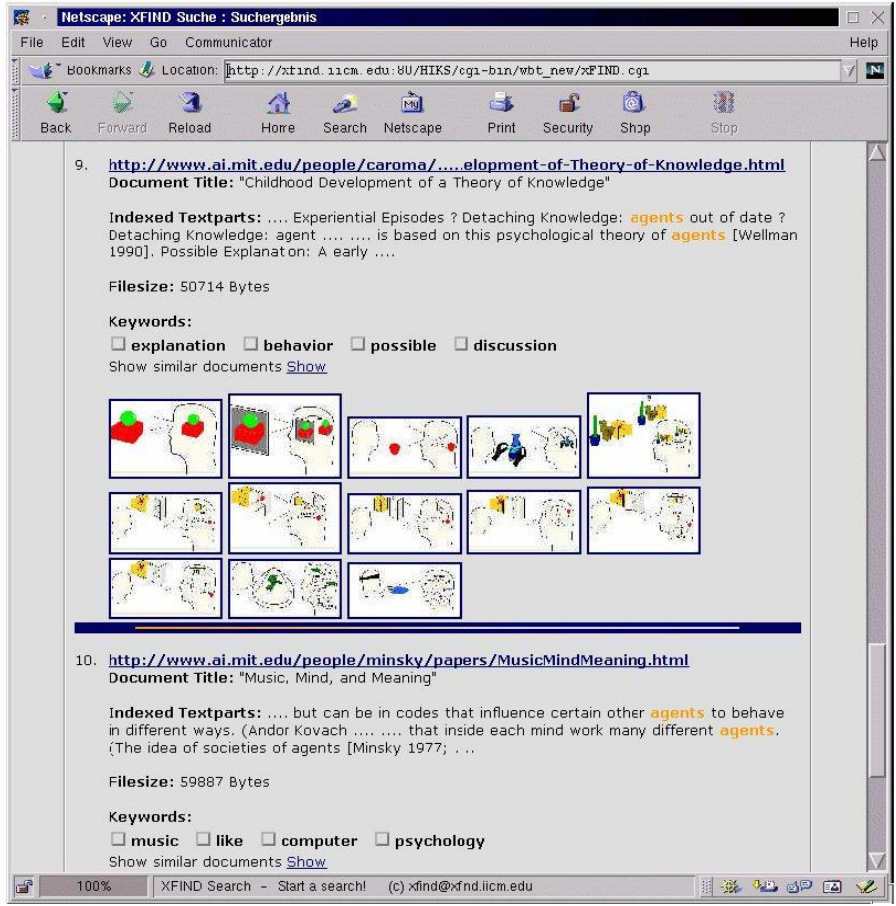

Figure 12: Thumbnails of figures extracted from a document in the xFind system (K. Andrews, Gutl, Moser, Sabol, & Lackner, 2001).

Term Distribution and Frequency

Thesauri, classifications and indexes are built to reduce the complexity of documents to single topics and hierarchies. Instead of reading the whole document these systems help to decide whether a document contains the information sought-after. Usually the information that a document is about a certain topic or certain topics are discussed in the document is binary information. It lacks information about how intense a topic is discussed or if it is the main topic or just a side reference being made. Term distributions and frequencies can indicate the importance of a topic in a document.

TileBars (M. A. Hearst, 1995) display the frequency and the distribution of query terms in a document. Documents are represented as rectangles. The length of the rectangle corresponds to the length of the document. Every segment of the document is represented by a square in the rectangle. If a set of query terms occur in a segment, its square is displayed in grey. The more often terms of one set occur, the darker the square is displayed, thus, indicating the terms' frequencies. Figure 13 shows query results with TileBars.

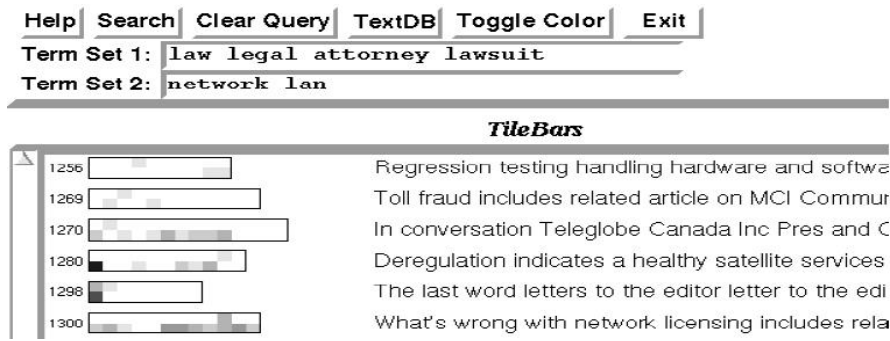

Figure 13: TileBars (M. A. Hearst, 1995) for two term sets on the left, document titles on the right.

TileBars were first proposed in 1995 (M. A. Hearst, 1995). Later studies have shown that even less complex versions of TileBars, only showing term frequencies by colour grade and not displaying their distribution inside the document, are an effective way to help the user in the search process and to improve his performance.

In the GridVis System (Weiss-Lijn, McDonnell, & James, 2001) the concept TileBar has been extended into an alternative interface to read documents. A metadata taxonomy is presented as a tree list in the first column. The taxonomy is structured hierarchically into topics and multiple levels of subtopics. In the second column a thumbnail of the document is presented horizontally above a grid. Each column of the grid corresponds to a paragraph of the document and each row to a topic. Due to the TileBar approach each cell indicates frequency and distribution of a topic in a paragraph by a colour code. In the third column the document itself is displayed. By clicking on a cell the corresponding paragraph is displayed and all paragraphs tagged with the selected topic are highlighted.

In a study (Weiss-Lijn, McDonnell, & James, 2002) in which users were asked to assess the relevance of a paragraph in certain documents on specific topics, no improvements in overall performance could be detected for participants using the GridVis system. The qualitative analysis revealed that the lack of familiarity with the system could be one of the main problems. The authors emphasized that a long-term study should be conducted to overcome this problem.

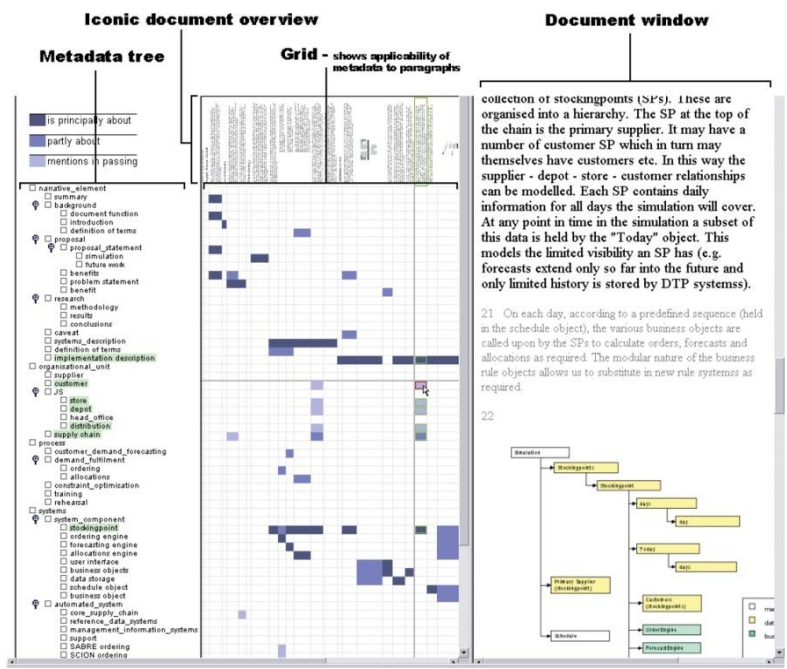

Figure 14: The GridVis system (Weiss-Lijn, McDonnell, & James, 2001).

For the VIDL System two ways of visualizing term frequencies of index terms have been developed and compared. These visualizations extend the index of a book with a visual representation of the content that lies behind an indexed term. In both visualizations the number of occurrences and the relation between the appearance on single pages and in page ranges can be displayed for every index term. They differ only in the way the visualization is done. In usability tests the authors only compared one of the visualizations against a text-only interface. The tests only assessed the participants' opinions on the system and not their retrieval performance. With respect to the document view, most of the participants were in favour of the system.

## Conclusion and Outlook

In twenty years of research a variety of approaches has been proposed and many prototypes have been developed to support the user of DLs in his search process on a visual level. For every part of the search process various techniques exist to support the user. But most of these techniques have not found their way into today's DLs. Nearly all prototypes have been discontinued. Most ideas have not been evaluated more than once in a relatively small study.

The main question for us is: Why have most of the results not been adapted into today's DLs? One simple answer could be that this is typical in Not all results of scientific research are supposed to be commercially beneficial or adaptable in large scale live environments. Looking at the different techniques and the studies that have been conducted we cannot say that the answer is that simple.

Taking a closer look at the results of the studies, we can see that usually quantitative results on the task performance and the accuracy of participants in a visual-based system are comparably poor or at least equally good as a strictly text-based system. On the other side, in questionnaires, the participants' opinions on the same visual-based system were positive and in favour of the system. Furthermore, not two evaluations have been conducted in the same way. Obviously it is not clear yet how to assess the usability and the quality of a visual-based search interface.

Another statement that all studies share is that the explanation for poor evaluation results is the users' lack of familiarity with the system. In many cases a completely new way of dealing with documents, result lists or queries is introduced and the user's task performance is measured against the *de facto* standard, a text-based search interface. On an abstract level this approach seems to be reasonable and is used analogical in other fields like Information Retrieval. In general, researchers in Information Visualization have to take into account that user performances are influenced by their experiences. And this influence has to be measured and included into an evaluation. E.g. after major release changes of software versions, the user performances often decrease in the beginning and the advantages of the software change are not exploitable immediately.

Also the difference between a visualization technique like TileBars or thumbnails and a visual-based prototype like xFind or INVISQUE is a problem. The prototypes often consist of multiple techniques running in parallel but in the study it is not assessed how results change by turning specific features on or off. In this way it is not possible to distinguish which techniques can be used successfully with others and in which combination.

To overcome these three main problems, a standard for evaluating visual-based search interfaces should be developed. The standard should define pro-cedures of how to conduct the study, which values should be assessed, contain long-term analyses in multiple steps and gauge the training curve of users over the study. Moreover, long term evaluations seem to be a crucial point. A good assess user performances over a long time period could be the logging of user behaviour in an actual DL. Only if a system is used by users with a real information need, it can be shown if certain visualizations are an added-value.

References

Andrews, K., Gutl, C., Moser, J., Sabol, V., & Lackner, W. (2001). Search result visualisation with xFIND. In Proceedings of the Second International Workshop on User Interfaces to Data Intensive Systems (UIDIS'01) (pp. 50–58). IEEE Computer Society Press. doi:10.1109/UIDIS.2001.929925

Andrews, Keith, Kienreich, W., Sabol, V., Becker, J., Droschl, G., Kappe, F., … Tochtermann, K. (2002). The InfoSky visual explorer: exploiting hierarchical structure and document similarities. Information Visualization, 1(3-4), 166–181. doi:10.1057/palgrave.ivs.9500023

Börner, K., Feng, Y., & McMahon, T. (2002). Collaborative visual interfaces to digital libraries. In Proceedings of the 2nd ACM/IEEE-CS joint conference on Digital libraries (pp. 279–280). New York, NY, USA: ACM. doi:10.1145/544220.544281

Chalmers, M., & Chitson, P. (1992). Bead: explorations in information visualization (pp. 330–337). ACM Press. doi:10.1145/133160.133215

Cugini, J., Laskowski, S., & Sebrechts, M. (n.d.). NIRVE paper. Design of 3-D Visualization of Search Results: Evolution and Evaluation. Retrieved February 9, 2013, from http://zing.ncsl.nist.gov/cugini/uicd/nirve-paper.html

Fox, E. A., Gonçalves, M. A., & Shen, R. (2012). Theoretical Foundations for Digital Libraries: The 5S (Societies, Scenarios, Spaces, Structures, Streams) Approach. Synthesis Lectures on Information Concepts, Retrieval, and Services, 4(2), 1–180. doi:10.2200/S00434ED1V01Y201207ICR022

Fox, E. A., Hix, D., Nowell, L. T., Brueni, D. J., Wake, W. C., Heath, L. S., & Rao, D. (1993). Users, user interfaces, and objects: Envision, a digital library. Journal of the American Society for Information Science, 44(8), 480–491. doi:10.1002/(SICI)1097-4571(199309)44:8<480::AID-ASI7>3.0.CO;2-B

Granitzer, M., Kienreich, W., Sabol, V., Andrews, K., & Klieber, W. (2004). Evaluating a System for Interactive Exploration of Large, Hierarchically Structured Document Repositories. In Proc. IEEE Symposium on Information Visualization INFOVIS 2004 (pp. 127–134). IEEE. doi:10.1109/INFVIS.2004.19

Hearst, M. (2009). Search user interfaces. Cambridge ; New York: Cambridge University Press.

Hearst, M. A. (1995). TileBars: visualization of term distribution information in full text information access. In Proceedings of the SIGCHI Conference on Human Factors in Computing Systems (pp. 59–66). ACM Press/Addison-Wesley Publishing Co. doi:10.1145/223904.223912

Hearst, M. A., & Karadi, C. (1997). Cat-a-Cone: an interactive interface for specifying searches and viewing retrieval results using a large category hierarchy. ACM SIGIR Forum, 31(SI), 246–255. doi:10.1145/258525.258582

Hienert, D., Schaer, P., Schaible, J., & Mayr, P. (2011). A Novel Combined Term Suggestion Service for Domain-Specific Digital Libraries. In S. Gradmann, F. Borri, C. Meghini, & H. Schuldt (Eds.), Research and Advanced Technology for Digital Libraries (Vol. 6966, pp. 192–203). Berlin, Heidelberg: Springer Berlin Heidelberg.

Jones, S. (1998). Graphical query specification and dynamic result previews for a digital library. In Proceedings of the 11th annual ACM symposium on User interface software and technology (pp. 143–151). Presented at the Proceedings of the 11th annual ACM symposium on User interface software and technology, ACM Press. doi:10.1145/288392.288595

Jones, S., McInnes, S., & Staveley, M. S. (1999). A graphical user interface for Boolean query specification. International Journal on Digital Libraries, 2(2-3), 207–223. doi:10.1007/s007990050048

Kim, B., Scott, J., & Kim, S. E. (2011). Exploring Digital Libraries through Visual Interfaces. Digital Libraries - Methods and Applications, 123–136.

Lam, H., & Baudisch, P. (2005). Summary thumbnails: readable overviews for small screen web browsers. In Proceedings of the SIGCHI Conference on Human Factors in Computing Systems (pp. 681–690). ACM Press. doi:10.1145/1054972.1055066

Morel, C. M., Serruya, S. J., Penna, G. O., & Guimarães, R. (2009). Co-authorship Network Analysis: A Powerful Tool for Strategic Planning of Research, Development and Capacity Building Programs on Neglected Diseases. (M. Tanner, Ed.)PLoS Neglected Tropical Diseases, 3(8), e501. doi:10.1371/journal.pntd.0000501

Robertson, G. G., Mackinlay, J. D., & Card, S. K. (1991). Cone Trees: animated 3D visualizations of hierarchical information. In Proceedings of the SIGCHI Conference on Human Factors in Computing Systems (pp. 189–194). ACM Press. doi:10.1145/108844.108883

Schatz, B. R., Johnson, E. H., Cochrane, P. A., & Chen, H. (1996). Interactive term suggestion for users of digital libraries: using subject thesauri and co-occurrence lists for information retrieval. In Proceedings of the first ACM international conference on Digital libraries (pp. 126–133). ACM Press. doi:10.1145/226931.226956

Sebrechts, M. M., Cugini, J. V., Laskowski, S. J., Vasilakis, J., & Miller, M. S. (1999). Visualization of search results. In Proceedings of the 22nd annual international ACM SIGIR conference on Research and development in information retrieval (pp. 3–10). Berkeley, California, USA: ACM Press. doi:10.1145/312624.312634

Shneiderman, B., Feldman, D., Rose, A., & Grau, X. F. (2000). Visualizing digital library search results with categorical and hierarchical axes (pp. 57–66). ACM Press. doi:10.1145/336597.336637

Weiss-Lijn, M., McDonnell, J., & James, L. (2001). Supporting Document Use Through Interactive Visualization of Metadata. In Proceedings of the ACM/IEEE Joint Conference on Digital Libraries.

Weiss-Lijn, M., McDonnell, J. T., & James, L. (2002). An Empirical Evaluation of the Interactive Visualization of Metadata to Support Document Use. In K. Börner & C. Chen (Eds.), Visual Interfaces to Digital Libraries (Vol. 2539, pp. 50–64). Berlin, Heidelberg: Springer Berlin Heidelberg.

Witten, I. H., Bainbridge, D., & Nichols, D. M. (2009). How to Build a Digital Library (2nd ed.). Burlington: Morgan Kaufmann.

Wong, W., Chen, R., Kodagoda, N., Rooney, C., & Xu, K. (2011). IN-VISQUE: intuitive information exploration through interactive visualization. In CHI '11 Extended Abstracts on Human Factors in Computing Systems (p. 311). ACM Press. doi:10.1145/1979742.1979720

Woodruff, A., Faulring, A., Rosenholtz, R., Morrsion, J., & Pirolli, P. (2001). Using thumbnails to search the Web. In CHI '01: Proceedings of the SIGCHI conference on Human factors in computing systems (pp. 198–205). ACM Press. doi:10.1145/365024.365098